\documentclass[8pt,british]{extarticle}
\usepackage[utf8]{inputenc}
\usepackage{geometry}
\usepackage{verbatim}
\usepackage{float}
\usepackage{mathrsfs}
\usepackage{amssymb}
\usepackage{graphicx}
\usepackage{setspace}
\makeatletter

\DeclareTextSymbolDefault{\textquotedbl}{T1}
\floatstyle{ruled}
\newfloat{algorithm}{tbp}{loa}
\providecommand{\algorithmname}{Algorithm}
\floatname{algorithm}{\protect\algorithmname}

\usepackage{fancyvrb}
\usepackage{url}
\usepackage{hyperref} 

\usepackage{listings}
\makeatother

\usepackage{babel}
\begin{document}
\title{Optimal design of experiments in the context of machine-learning inter-atomic
potentials: improving the efficiency and transferability of kernel
based methods. }
\author{Bartosz Barzdajn and Christopher P. Race}
\maketitle
\begin{abstract}
Data-driven, machine learning (ML) models of atomistic interactions
are often based on flexible and non-physical functions that can relate
nuanced aspects of atomic arrangements into predictions of energies
and forces. As a result, these potentials are as good as the training
data (usually results of so-called \emph{ab initio} simulations) and
we need to make sure that we have enough information for a model to
become sufficiently accurate, reliable and transferable. The main
challenge stems from the fact that descriptors of chemical environments
are often sparse high-dimensional objects without a well-defined continuous
metric. Therefore, it is rather unlikely that any \emph{ad hoc }method
of choosing training examples will be indiscriminate, and it will
be easy to fall into the trap of confirmation bias, where the same
narrow and biased sampling is used to generate train- and test- sets.
We will demonstrate that classical concepts of statistical planning
of experiments and optimal design can help to mitigate such problems
at a relatively low computational cost. The key feature of the method
we will investigate is that they allow us to assess the informativeness
of data (how much we can improve the model by adding/swapping a training
example) and verify if the training is feasible with the current set
before obtaining any reference energies and forces -- a so-called
off-line approach. In other words, we are focusing on an approach
that is easy to implement and doesn't require sophisticated frameworks
that involve automated access to high-performance computational (HPC).\emph{
}%
\end{abstract}

\section{Introduction}

Inter-atomic potentials are surrogate models replacing complex quantum-mechanical
(QM) calculations with fast-to-evaluate functions, directly or indirectly
dependent on the positions of atoms, returning forces and energies.
With such potentials, we can simulate millions of atoms at timescales
of nanoseconds or microseconds; something beyond the reach of QM models. 

\begin{figure}
\includegraphics[width=1\textwidth]{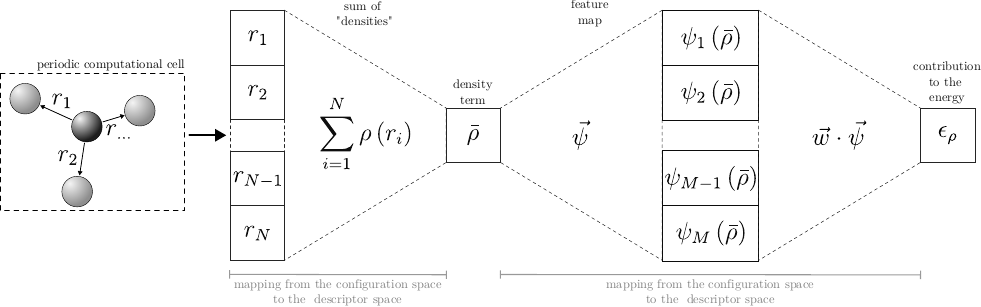}

\caption{Illustration of the problem of double mapping using the example of
the EAM model. Here we focus only on the contribution of a specific
atom to the total energy and consider only the embedding representing
the many-body interactions. The density $\bar{\rho}$ in principle
refers to the local electronic density and consists of contributions
$\rho\left(r_{i}\right)$ from neighbouring atoms, where $\rho$ is
a non-linear function that can also depend on adjustable parameters,
while $r_{i}$ represents the distance to a specific neighbour. We
also assume that considering the first N neighbours provides almost
complete information. The quantity $\bar{\rho}$ can be considered
as a 1-dimensional descriptor of the local environment. The contribution
to the energy will be a non-linear function of this quantity. However,
in this example we assume that it can be expressed in a linear basis
to illustrate the model complexity represented by the number of features
M. An example of a feature can be $\bar{\rho}$ raised to the \emph{k}-th
power in the polynomial representation of the mapping.}
\label{fig:double_map}
\end{figure}

When regarded as a regression problem, the development of these potentials
is challenging. Firstly, the problem cannot be easily formulated using
a space with a comprehensive metric (e.g. $\mathbb{R}^{3\times N}$,
$N$ being a number of atoms). A useful potential needs to be applicable
to a variety of configurations requiring different numbers of atoms
in a computational cell. Hence, the map, linking to energies and forces,
will be defined (implicitly) on a collection of positions of various
sizes. For this reason alone, formulation of potentials will be facilitated
by descriptors of chemical environments -- functions that represent
a collection of atoms as a vector, tensor or a set (with the simplest
form being a list of pairwise distances). As a result, we have two,
rather than one, consecutive and non-trivial relationships. The first
maps the positions of atoms to their abstract representatives and
is defined by our decision to use a particular descriptor. The second
is defined by a function that accepts a descriptor as an input and
outputs energies and/or forces. This has been illustrated on the example
of the embedded-atom model (EAM) (see e.g. A.F. Voter in \cite{Westbrook1994-kv,Voter_EmbeddedAtom__})
in fig \ref{fig:double_map}.

For example, when we wish to train an inter-atomic potential, we need
to choose a suitable descriptor and determine the right model complexity.
This way, we can maximise the precision of the fit without introducing
a bias. However, it is easy to make a mistake by selecting a simple
descriptor that cannot differentiate between many distinct chemical
environments. At the same time, we can define a very flexible regression
model that can accommodate all the differences in the training data.
As a result, we will encounter the problem of over-fitting despite
defining a model with the right overall capacity. This is not an unusual
problem in development of EAM potentials. 

In response to such challenges, researchers are turning to machine-learning
methods, such as, after \cite{Kocer_Neural_ARPC_2022}, Neural Network
Potentials \cite{Lorenz_Representing_CPL_2004}, kernel based methods
like the GAP -- the framework for `atomistic' Gaussian process
regression \cite{Bartok_Gaussian_PRL_2010} and Gradient-domain Machine
Learning models \cite{Chmiela_Exact_NC_2018}, Moment Tensor Potentials
(MTP) \cite{doi:10.1137/15M1054183,Shapeev_Moment_MMS_2016}, methods
with a physically inspired basis like Atomic Cluster Expansion (ACE)
\cite{Drautz_Atomic_PRB_2019}, and many others \cite{Mishin_Machinelearning_AM_2021,Friederich_Machinelearned_NM_2021}. 

In this paper we focus on methods that are coupled with extensive,
or rather `expressive', descriptors, such as the SOAP (smooth overlap
of atomic positions, \cite{Bartok_Representing_PRB_2013}), designed
to be applicable to many materials, distinguish between alloying elements
and reflect nuanced many-body interactions. In other words, models
flexible enough to be as good as the data provided. The price we will
have to pay for high levels of accuracy is susceptibility to insufficient
information. Once we decide to use descriptors and regression models
that are sufficiently complex to be error-free and universal, we need
to make sure that the training data is sufficient as well. This way
interpolations and extrapolations will not fail as soon as we try
to make predictions on examples moderately distinct from the training
set. 

It is deceptively tempting to manage this problem using brute force
and integrate into the training set as many examples as possible.
However, it can be inefficient, unreliable, and even with access to
high-performance computing (HPC) facilities prohibitively expensive. 

The choice of training examples are often motivated by physics and
intuition. But what is a distinct training example from the perspective
of a researcher, might not be distinguishable by the model or descriptor.
What is informative may not be representative. For instance, a training
set might be improved significantly by the inclusion of unrealistic
configurations that provide better coverage of possible descriptor
values. Likewise, polluting a data set with contradicting (\emph{e.g.
}when descriptor cannot distinguish between physically different examples)
or irrelevant information might result in a loss of accuracy or performance
respectively. 

We need a systematic approach, as we have to navigate an extensive
set of possible atomic configurations, through at least two maps (configuration
-- descriptor and descriptor -- predictions of energies, etc.),
and most likely without a metric. It is not a surprise that the development
of a potential is such an extensive and risky task.

In practice, researchers often resort to using some form of active
learning, like the classical Cs\'{a}nyi et al. learn-on-the-fly approach
\cite{Csanyi_Learn_PRL_2004}, where model, or models, trained on
partial training sets are used to query if new candidates are a good
addition to the current set. As reported by Jinnouchi et al. in their
overview, these strategies can result in a significant reduction in
the time required for training \cite{Jinnouchi_OntheFly_JPCL_2020}.
Albeit, they still require access to vast computational resources
and complex software infrastructure. Given that the training and assessment
will likely involve shared HPC resources, with strict management policies,
an infrastructure that might be difficult to implement. Furthermore,
initial queries are made using models that can be significantly flawed
unless the original database was already extensive at the beginning.
Which brings us right back to the initial point.

Candidate searches can be more efficient if they can be done without
labelling, \emph{i.e.} obtaining values for associated quantities
of interest, which usually means estimating energies and forces with
ab initio QM calculations. With kernel methods, for example, we can
use the pivoted low-rank approximations to mark a subset of candidates
for labelling. After all, the idea behind this approximation is to
select a subset of data that is most representative of the whole,
\emph{i.e.} giving the closest approximation to the full kernel matrix.
This feature of labelling is even included in the original implementation
of the GAP \cite{Bartok_Gaussian_PRL_2010}. Likewise, as shown by
Podryabinkin and Shapeev in their active learning scheme \cite{Podryabinkin_Active_CMS_2017},
or more recently by Lysogorskiy et al. \cite{Lysogorskiy_Active_PRM_2023},
we can apply in the context of training of potentials criteria developed
for statistical design of experiments, i.e. using the rigorous language
of statistics to decide in advance what information to add or start
with.

We argue that classical concepts of statistical planning, such as
optimal design of experiments \cite{Rasch_PWN_1991}, many of which
were developed when even supercomputers could not match the speed
of today's smartphones, can indeed provide a basis for efficient solutions
to the problems outlined above. Optimality in this context means that
for a given number of training examples (budget), we can choose those
that minimise the uncertainty associated with the model parameters.
Optimal designs are unique to linear basis function models and thus
to kernel methods such as GAP \cite{Rasch_PWN_1991}. Associated methods
allow us to manage model uncertainty and verify the feasibility of
learning before even the first calculation is made. As such, they
allow rapid exploration and sifting through a large number of training
examples. At the same time, they provide access to well-established
measures of quality that are independent of labels/outputs. These
measures are one of the main advantages of statistical planning, and
a certain advantage over `vanilla' active learning.

In our work, we will present how to optimise data for on kernel-based
methods rather than weight-space models like ACE or MTP, as the associated
optimality criteria and algorithms are less known and more difficult
to implement. For the reasons outlined above, we also aim for solutions
that provide optimal, or more realistically optimised, sets without
the need for retraining and estimation of empirical errors. In this
respect, our work is related to that of Karabin and Perez \cite{Karabin_Entropymaximization_JCP_2020}.
However, we focus on methods that aim to find optimal solutions by
directly considering optimality criteria rooted in statistical theory.

\section{Methodology}

To emphasise the key characteristic of designs we are focusing on,
we refer to them as a priori designs. As mentioned before, an optimal
training set can be constructed before any ab initio calculation or
experiment is made. The key is to choose an appropriate measure. For
example, in linear regression with constant normally distributed ‘noise’,
the variance-covariance matrix of regression coefficients is $\Sigma=\sigma^{2}\left(X^{\top}X\right)^{-1}$,
where $\sigma^{2}$ is the variance of the `noise', and $X$ is
the design matrix -- our usual assembly of inputs associated with
features. It immediately transpires that by selecting appropriate
inputs to $X$, we can minimise $\Sigma$ and we do not need any knowledge
about the $\sigma^{2}$ (often it is a const. value) or outputs (the
$y$ 's) in our data set. More details can be found in appendix \ref{sec:the-concept}.

There are many well-established criteria for optimality of designs
that are more suitable than the direct optimisation of $\Sigma$.
The most common is the \emph{D} -- optimality that seeks to maximise
the determinant of the observed Fisher information matrix, which is
given by $\sigma^{-2}X^{\top}X$ for linear models with the constant
`noise' (notice the similarity with the formulation of $\Sigma$).
This particular criterion was used in earlier work by Podryabinkin
and Shapeev to mark candidates for evaluation/labelling \cite{Podryabinkin_Active_CMS_2017}.
However, we have to keep in mind, that each family of models, be it
ordinary least-squares or kernelised ridge regression (KRR, \cite{Murphy_Machine__2012}),
will require different strategies and criteria of optimality. Moreover,
non-linear models will have only local optimal designs for a specific
range of parameters \cite{Rasch_PWN_1991}.

In our work, we focus on the GAP model framework of Gaussian process
regression (GPR). This framework can be categorised as state-of-the-art
atomistic modelling that can give predictions almost as good as the
training data. 

To reiterate argument from the introduction, high-quality data, usually
in the form of atomic configurations and associated energies and forces,
are expensive to generate, even with modern hardware and efficient
implementations of the density functional theory (DFT) \cite{Kohn_SelfConsistent_PR_1965}.
Furthermore, we would like to use the model to make predictions on
extensive computational cells that are beyond the reach of the DFT
method rather than interpolate a large number of calculations. Hence,
we cannot test the model by comparing it with a reference, nor can
we rely on the model prediction variance as it is unreliable when
it comes to highlighting errors in the definition of the descriptor.
All this renders any form of active learning a less appealing choice.

To address design criteria for GPR, and as such GAP, we need to refer
to the fundamentals. A Gaussian process (GP) is a collection of random
variables with joint Gaussian distribution
\[
f\left(\vec{x}\right)\sim\mathcal{GP}\left(m\left(\vec{x}\right),k\left(\vec{x}\right)\right),
\]
specified by mean function $m$ and covariance function $k$. The
model, specified by rules to evaluate $m$ and $k$, is defined by
the posterior distribution conditioned on the data (\cite{Rasmussen_Gaussian__2006},
chapter 2). In the GPR framework, each element of the training set
and each extrapolation point corresponds to a degree of freedom of
a multivariate Gaussian distribution. The expectation and variance
of this distribution are defined by a kernel matrix - a matrix of
inner products between all data points (training and predictions).
Formulation of GPR and the framework in the context of atomistic modelling
can be found in \cite{Bartok_Gaussian_IJQC_2015} and \cite{Deringer_Gaussian_CR_2021}. 

The GPR can be regarded as a linear method with respect to weights
(model parameters) if we choose to formulate it in the so-called weight-space
view. In the function-space view, it is strongly related to KRR\footnote{Kernel methods replace the explicit evaluation of features with their
dot products, which we can calculate using fast-to-evaluate formulas
-- kernel functions. This way, we can implicitly work with a large,
or infinite, number of features. However, full matrices used in regression
will be as large as the dataset.}. Both methods share the same estimator of expectation if the penalty
corresponds to the prior variance (data `noise').

Relationships between these methods suggest that optimal designs are
to some extent transferable. However, we need to be careful. In our
experiments, when we tried to minimise the maximum prediction variance
of the KRR model in an active learning framework, we created a dataset
that was performing worse, when used to train a GAP model, than the
source-sampling. Here, the source-sampling is a method that is used
to generate a large number of candidates, that are later reduced by
the algorithm to create an optimised training set.

The most straightforward design for GPR is the maximum entropy (MaxEnt)
sampling introduced by Shewry and Wynn in \cite{Shewry_Maximum_JoAS_1987}.
We refer here to the entropy defined as the expectation of information
content, also known as the Shannon information. In principle, it is
a different quantity the entropy defined in physics. The optimality
criterion authors propose (section 4) is 
\begin{equation}
\max\log\det\mathbf{K},\label{eq:max_ent_gp}
\end{equation}
where $\mathbf{K}$ is the covariance matrix of the multivariate Gaussian.
This is the part of the covariance matrix associated with the training
data in the GPR framework.

As discussed by the authors, this criterion attempts to maximise the
variability in the training set. In other words, their distinctiveness
and the coverage of the domain. Initially, we used this criterion
in combination with a modified exchange algorithm \cite{Miller_Algorithm_AS_1994},
i.e. an algorithm that swaps candidates in the training set with the
pool of potential replacements until the set becomes an optimal representation
of the domain. However, we were concerned with its low efficiency
of exploration as the algorithm required, to be computationally efficient,
generating a covariance matrix for all the data, limiting the number
of candidates we can consider at the same time.

\begin{figure}
\begin{centering}
\includegraphics[width=1\textwidth]{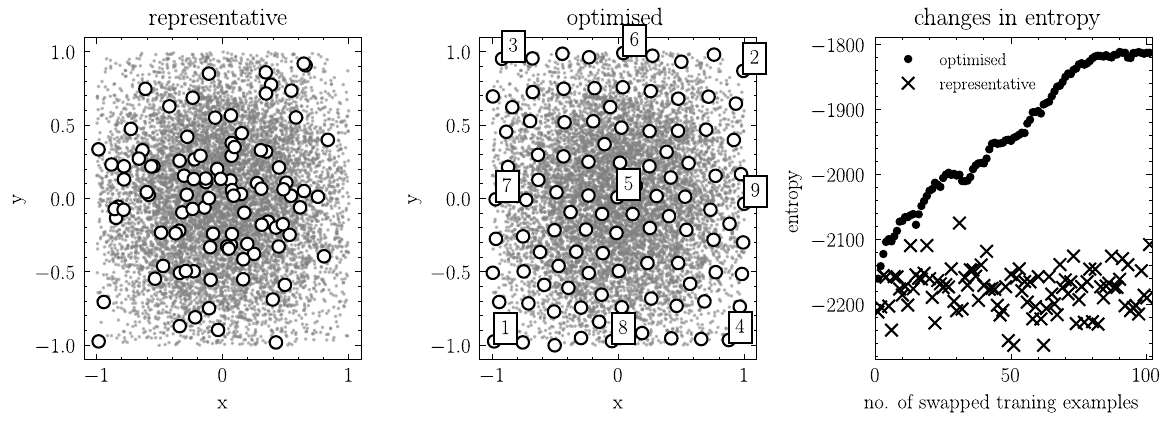}
\par\end{centering}
\caption{Application of the Golchi and Loeppky algorithm \cite{Golchi_Monte__2016}
for max-min designs. In this example, we are optimising the training
set for Gaussian process regression. The distance measure follows
the definition \ref{eq:k_dist} and is based on a Gaussian kernel
with unit scale parameter. The training points are selected from a
pool of candidates generated using strongly biased sampling. This
pool consist of $10^{4}$ samples from the normal distribution $\mathcal{N}\left(0,0.5\right)$.
On the plots they are represented by small grey points with opacity.
Such an example simulates conditions under which we have to select
atomistic configuration for training of ML potentials. The first plot
on the left illustrates a representative design, which is a random
sub-sample of 100 candidates. This plot also illustrates how biased
the underlying sampling is. Next, in the middle, we show the distribution
in the optimised training set, also consisting of 100 examples. The
index indicates the order of addition of the first 9 optimal points,
revealing the algorithm's strategy of filling empty spaces after enclosing
a domain. Finally, the plot on the right shows how the entropy changes
when we replace the random/representative examples with the optimised
ones. Entropy for representative sets corresponds to different realisations
of this design. Here, the entropy of the resulting kernel matrix $K$
is proportional to $\log\left(\det K\right)$ \cite{Shewry_Maximum_JoAS_1987}.
Note that the aim is not to provide a uniform importance sampling,
but optimal training points for a given regression method. In this
case, however, these objectives coincide so that we can visually assess
the quality of the solution. Other models may have a different solution.
See the example a) in figure \ref{fig:design_example}. Finally, this
results can be easily recreated using the alghorithm \ref{alg:opt_grid}
from appendix\textbf{ }\ref{sec:maxmin_python}. We only need to replace
the }

\label{fig:alg_test}
\end{figure}

For this reason, we decided to apply a conditional max-min design
with an algorithm presented in \cite{Golchi_Monte__2016} that performed
well in comparison with our previous solutions. This design aims to
maximise the minimum distance between samples. The algorithm is illustrated
in figure {[}fig:the\_alg{]}. A simple and efficient implementation
of this algorithm can be found in the appendix \ref{sec:maxmin_python},
as well as in \cite{barzdajn_2024_11191242} along with its applications
in other contexts. 
\begin{figure}
\begin{centering}
\includegraphics[width=0.85\textwidth]{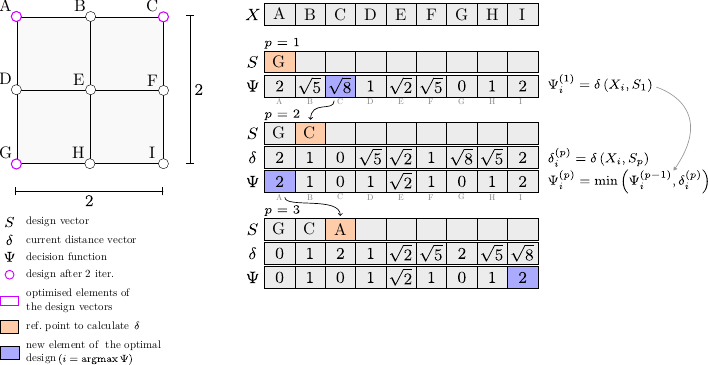}
\par\end{centering}
\caption{Illustration of the Golchi and Loeppky algorithm \cite{Golchi_Monte__2016}
for max-min designs, which maximises the minimum distance, using the
example of nine points on a square grid. The state in each iteration
is defined by the design vector $S$ and vector $\Psi$. The vector
$\Psi$ can be considered as a 'decision' function while $X$ represents
the pool of candidates. Initially, $\Psi$ consists of distances between
the first element and the remaining elements. In the following iterations,
$\Psi$ is updated element-wise according to $\Psi_{i}^{(p)}=\min\left(\Psi_{i}^{(p-1)},\delta_{i}^{(p)}\right)$,
where p is the iteration index, $\delta_{i}^{(p)}$ is the distance
between $p\mathrm{-th}$ and $i\mathrm{-th}$ candidate and $\delta\left(\cdot\right)$
is the distance function. In each iteration we find the maximum value
of update $\Psi$. Position of this element indicates the new optimal
design point. }
\label{fig:the_alg}
\end{figure}
 It proceeds in a greedy manner, \emph{i.e.} candidates are ordered
in terms of importance, starting with the most informative examples,
in other words the most distanced. As the distance, we choose the
squared kernel distance (\cite{Phillips_Gentle__2011}) 
\begin{equation}
D_{K}^{2}=k\left(p,p\right)+k\left(q,q\right)-2k\left(p,q\right),\label{eq:k_dist}
\end{equation}
where $K$ is the covariance function between configurations $p$
and $q$. We chose a measure based on the similarity kernel in order
to create a design that is directly related to the regression method.
Here, we used a complete covariance, as described in \cite{Bartok_Gaussian_IJQC_2015},
consisting of radial-basis functions for pair-potentials and a polynomial
kernel with the SOAP descriptor for many-body interactions. Defining
the distance in such a way also i consistent with optimality criterion
\ref{eq:max_ent_gp} (see figure \ref{fig:alg_test}).

According to our experiments with the Euclidean metric and highly
biased sampling, the algorithm generates space-filling designs that
embrace the whole domain. For linear least-squares regression, such
designs also tend to minimise maximum prediction variance, and as
such, they are consistent with the aims of G--optimality \cite{Johnson_Minimax_JoSPaI_1990,Mullera_Spacefilling_PES_2011}.
Figure \ref{fig:alg_test} demonstrates the test of the algorithm
on a simple example of two-dimensional Cartesian space. 

The distance is the same as in \ref{eq:k_dist}. However, in this
example we selected the Gaussian kernel. As such, the solution is
an optimised set with respect to the performance of the GPR regression,
rather than optimal filling of the space. Although, for a Gaussian
kernel with a large scale parameter these goals will coincide. 

\section{Numerical (in silico) experiment }

It is reasonable to begin the development of a training database for
GAP with a collection of examples that will inform the model about
elastic deformations (see e.g. \cite{Szlachta_Accuracy_PRB_2014}).
This first step will serve as our case study. In other words, we wish
to create a database of elastically deformed structures such that
we can predict the elastic energy density of a pure Zr lattice. Our
single training database will consist of 1000 examples and include
hcp, fcc and bcc structures.

To create an optimised training set, allowing us to train a more accurate
model, first, we generate an as extensive as possible set of candidates.
The core idea behind the optimisation is to sift through contenders.
Bear in mind that an empirical model is as good as the data it interpolates.
Hence, we need to make sure the space is covered appropriately and
that we have a sufficiently good source-sampling -- one that can
draw from every possibility, here defined by the structure and the
strain state, with a reasonable, but not necessarily optimal, probability. 

Each time, the transformation matrix, used to deform a randomly picked
structure, is generated from a random uniform vector of eigenvalues
(we essentially reversed the eigenvalue decomposition) and rotated
using Euler angles picked indiscriminately from the group of 3D rotations. 

Initially, we generated the deformation using uniform sampling of
the elements of the strain tensor. However, in our experiments this
sampling was severely under-performing. As such, optimised sets were
always significantly better and it was difficult to judge the results
against the `random' sampling. In addition, we could see the benefits
of the statistical planning right from the beginning as it forced
us to improve the framework.

In the case study we used a converged candidate set size of 100,000
from which we selected 1000 candidates using the framework we discussed
earlier, i.e. the conditional max-min design with the greedy algorithm
\cite{Golchi_Monte__2016}. %

As planned for the final atomistic model, the kernel matrix for the
data-set optimisation consisted of two components. We used the Gaussian
kernel to estimate pair-wise interactions, with a scaling factor of
0.8 and a length scale of 0.3, while for many body interactions, we
selected the SOAP descriptor and 4th-order polynomial with a 0.3 scaling
factor. The SOAP descriptor was initiated with the following string
defining the parameters: \texttt{soap cutoff=6.8 l\_max=10 n\_max=10
normalize=T atom\_sigma=0.15 n\_Z=1 Z=\{40\}}. The cut-off of the
pair-wise distance descriptor was set to 7.0. Parameters were selected
by attempting to minimise the number of overlapping (kernel distance
close to zero) and orthogonal (kernel distance close to 1) examples
in the trial candidate set. In other words, we selected a descriptor
that can distinguish between the most similar chemical environments
and, at the same time, that won't be too sensitive to changes in atomic
positions and lose the ability to quantify similarity of most distinct
configurations (clipping effect).

Reference data (labels for training sets) were evaluated using the
DFT method implemented in VASP \cite{Kresse_Efficiency_CMS_1996,Kresse_Efficient_PRB_1996,Kresse_Initio_PRB_1993,Kresse_Ultrasoft_PRB_1999}.
We choose the APW basis with an energy cut-off of 600 eV, automatic
determination of number of k-points (with 60 subdivisions) and Methfessel-Paxton
smearing method \cite{Methfessel_Highprecision_PRB_1989}.

For a given training set, having obtained energies, forces and virials,
we optimised kernel hyper-parameters $\theta$ (parameters of descriptors)
by maximising the log-likelihood:
\[
\mathcal{L}\left(\theta\right)=-\frac{1}{2}\left(y^{\top}Q^{-1}y+\ln\det(Q)+n\ln\left(2\pi\right)\right),
\]
where $Q=K+\sigma I$, $Q$ is the kernel matrix of the additive model
(pair-potential and the SOAP kernel) and $\sigma$ is the prior variance
of the data. In the above, $y$'s are DFT energies and the optimal
solution does depend on the results. Therefore, each data set will
have a separate set of associated optimal parameters.

Using the optimised training set and meta-parameters we trained a
GAP model using the \texttt{gap\_fit} program from the QUIP library
\cite{Bartok_Gaussian_PRL_2010}. Contrary to the optimisation of
hyper-parameters, the model fit was conducted using forces and virials
as well. This way we can asses the practicality of our approach. %

\section{Results and conclusions}

In this section, we will assess the methodology and the implementation
of the algorithm. We begin by investigating how the algorithm affects
the similarity of the training examples. 

\begin{figure}
\begin{centering}
\includegraphics[width=0.75\textwidth]{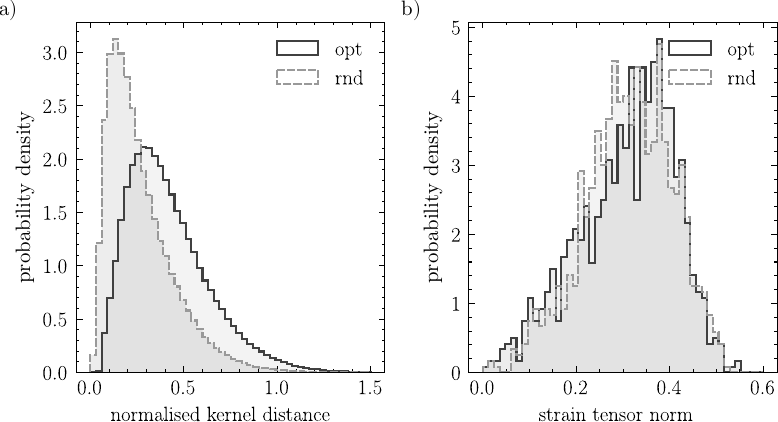}
\par\end{centering}
\caption{Comparison of pairwise distances within a random and an optimised
training set. Figure a) shows normalised probability histograms of
the kernel distance. Figure b) shows the distribution within a training
set of the $L_{2}$norm of the deformation tensor, which we use to
quantify its magnitude. The lack of a strong difference between the
locations of these distributions shows that the reduction in overlap
has not been achieved by simply pushing the \textquotedbl amount\textquotedbl{}
of deformation. }

\label{fig:similiarity_compare}
\end{figure}

We compare the randomly selected ('representative') deformed structures
with those from the optimised set. The results are shown in figure
\ref{fig:similiarity_compare}.

It is immediately apparent that the algorithm works by removing overlapping
examples and emphasising diversity, as indicated by the shift of the
entire distribution, including the minimum, towards larger distances.
Note that the opposite of overlap is when all examples are orthogonal
and the dissimilarity, i.e. the kernel distance, is 2 in our case,
as we have omitted the normalisation in equation \ref{eq:k_dist}.

Keep in mind that a successful optimisation is not always guaranteed.
When using high-dimensional descriptors and kernel-based distances,
the algorithm may favour solutions where all examples are orthogonal
to each other. This would mean that the training set is useless for
models that are supposed to be interpolators. We can fall into this
trap by selecting extremely deformed examples. Therefore, we had to
restrict the search space accordingly. In our case, we did this by
limiting the maximum eigenvalues of the deformation matrix. This highlights
the key chellenge of the presented methodology, which is the choice
of the distance or dissimilarity measure.

While we demonstrated that the methodology improves the quality of
a training set, we need to show that this change is significant in
a realistic scenario. To do this, we generated two optimised training
sets, OPT1 and OPT2, consisting of 1000 examples selected from two
independent pools of candidates (100,000 per pool). Additionally,
from each we selected randomly and indiscriminately a subset of 1000
candidates, creating two representative training sets: RND1 and RND2.
Here, OPT stands for ‘optimised’ and RND for ‘random’ or representative
of the source-sampling.

\begin{figure}
\begin{centering}
\includegraphics[width=1\textwidth]{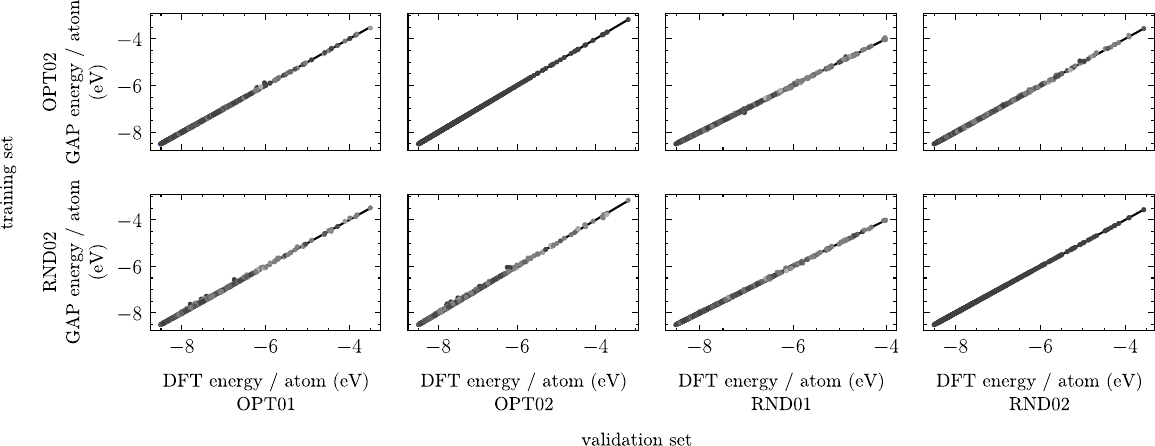}
\par\end{centering}
\caption{Examples of performance when predicting energy per atom on training
and validation sets. Brightness indicates prediction error from fig.
\ref{fig:cross_val}.\label{fig:examples_performance}}

\medskip{}

\begin{centering}
\includegraphics[width=1\textwidth]{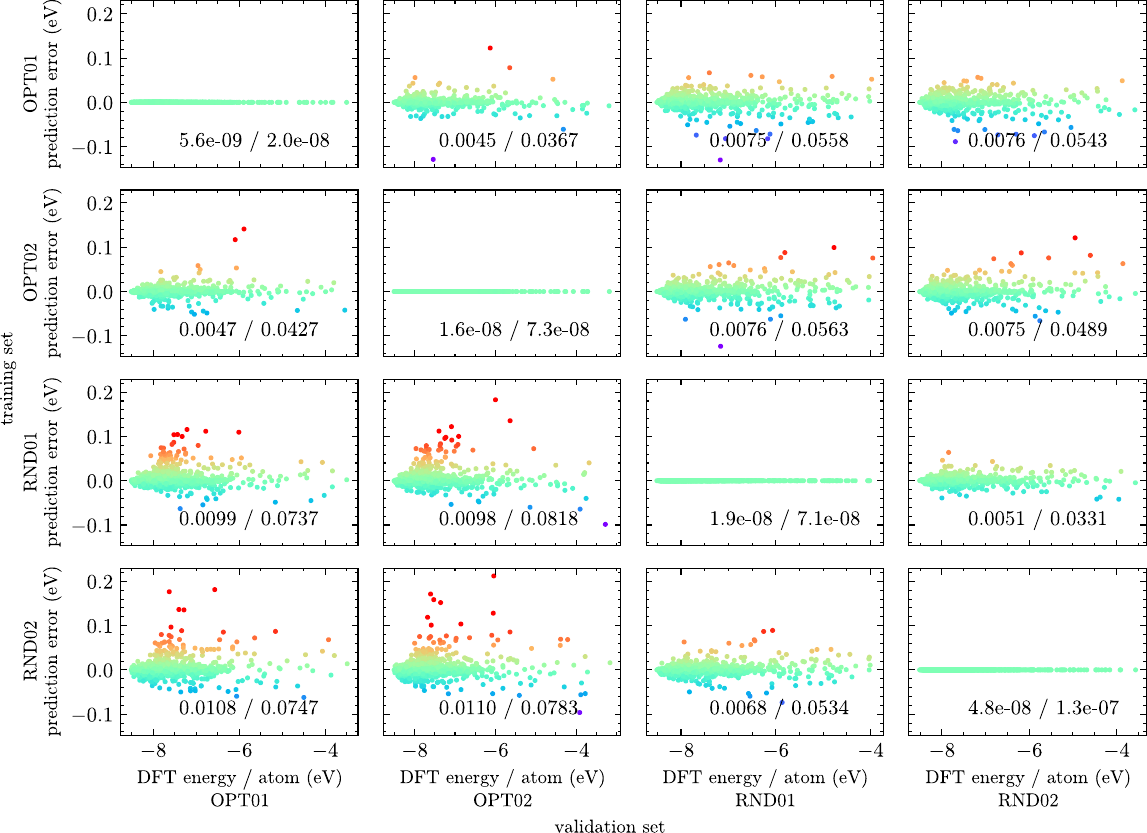}
\par\end{centering}
\caption{Results of cross-validation. Here (GAP) models are defined by their
corresponding training sets and organised in rows. Each column corresponds
to validation set on which model was tested. Plots on the diagonal
illustrate fitting performance. Colour indicates the position on the
y-axis. Summary statistics are: standard deviation (first number in
the label) of the absolute error (DFT results versus GAP prediction),
and 0.99 quantile (second number in the label). \label{fig:cross_val}}
\end{figure}

The idea is to gauge how much improvement the optimisation introduced
by comparing models trained on each set. The procedure is closely
related to the concept of cross-validation. In the assessment, we
utilised all DFT energies and GAP predictions made on all configurations.
Having four training sets and, as such, four models, we can make 16
comparisons in total. For example, we take the GAP model trained on
the OPT1, evaluate energies on configurations from OPT1, OPT2, RND1,
RND2 and compare GAP predictions with DFT references. When we test
the model predictions on its own training set, essentially, we evaluate
the goodness of fit. Otherwise, we are enquiring about the off-sample
performance. Results are presented in figure \ref{fig:examples_performance}
and \ref{fig:cross_val}. 

It immediately transpires that the optimised training sets delivered
a more consistent and overall better performance than their representative
(random) counterparts. Here, we measure the performance as the worst
off-sample performance given by the highest, among all test sets,
0.99 quantile of the absolute error. We demonstrated that we can achieve
better transferability in exchange for some level of precision in
a more narrow context. In other words, optimised sets are likely to
have a better coverage of the whole domain in exchange for fewer samples
around the mode of the source-sampling, i.e. the distribution to generate
the initial pools of candidates. It is exactly what we expect from
the optimal design. The argument is reinforced by the fact that models
trained on RND1 and RND2 sets did well when tested against each others
corresponding training sets. At the same time, when evaluated on OPT
sets, their performance was worse than the worst performance of OPT
models.

There are two main considerations that we have to keep in mind when
interpreting the results. First, the underlying generation of training
examples has already been improved as a result of initial analysis
in the context of statistical planning. However, it is not always
easy to improve the source-sampling as it was in the case of elastic
deformations of perfect lattices. Secondly, in the optimisation we
used a kernel matrix associated with energies. Yet the final training
sets also consisted of forces and virials as a part of associated
labels/outputs. Therefore, we have shown that even a simplified methodology
can provide quantifiable benefits in realistic scenarios.

In summary, we demonstrated that the optimal design and statistical
planning is an effective tool that can improve the reliability and
generalisability of data-driven models. The two main advantages over
the active learning are providing measures of quality that allow comparison
of training sets and the ability to improve them without the necessity
of evaluating labels, i.e. in this context, energies, forces and virials. 

At the same time, we demonstrated that empirical methods of performance
evaluation, such as the well-known test-train-split framework, should
be applied carefully. It is relatively easy to overestimate the predictive
power of a model and fall into the trap of confirmation bias. As we
saw when comparing representative sets. 

For example, when the method of creating training examples probes
only a narrow part of a domain, we obtain \textquotedbl dense\textquotedbl{}
sampling that can give us deceptively good interpolations in this
region. While it is not a problem in Euclidean spaces, with the high-dimensional
descriptors, it might be extremely difficult, or even impossible,
to define a metric that will allow us to sample the domain \textquotedbl uniformly\textquotedbl .
In other words, we always have to keep in mind that we are always
in danger of replicating biases introduced by the source-sampling. 

The statistical planning of experiments may help us to overcome these
challenges. However, only when certain conditions are satisfied. First,
we can select from a large enough pool of candidates. Secondly, we
have a method to adequately sample the whole domain.

The application of optimal design in the context of ML potentials
is still relatively unique. Hence, there are many areas for potential
development. First, we need ways to define sufficiently good underlying
sampling, especially for many body representations. We suggest that
developments related to exploring configuration space are aligning
with this goal and can be used to ensure appropriate coverage of the
descriptor space. Secondly, another essential part of the methodology
is the efficient optimisation algorithm. The conditional max-min design
or exchange algorithms can be adequate only when descriptors, or the
kernel matrix, are precalculated. Therefore, for large sets of candidates,
memory requirements can be limiting, and there is a need for efficient
lazy algorithms. 

\section{Acknowledgements}

We would like to kindly acknowledge The Engineering and Physical Sciences
Research Council (EPSRC) for funding the MIDAS project (Mechanistic
understanding of Irradiation Damage in fuel Assemblies -{}- ref. EP/S01702X/1).
C P Race was funded by a University Research Fellowship of the Royal
Society. Calculations were performed on a computational cluster, maintained
by the Computational Shared Facility, The University of Manchester.

\section{Author contributions}

\textbf{Bartosz Barzdajn}: Writing - Original draft, Conceptualization,
Methodology, Software, Validation, Formal analysis, Investigation,
Data curation, Visualization. \textbf{Christopher Race}: Writing -
Review \textbackslash\& Editing, Conceptualization, Methodology,
Validation, Formal analysis, Supervision, Project administration,
Funding acquisition. 

\bibliographystyle{unsrturl}
\bibliography{OPT_DES,MLIP,ML,QM,OPT_DES_MLIP}

\appendix

\section{Introducing the concept of the optimal design\label{sec:the-concept}}

We will illustrate the concept of optimal design using the simplest
regression model 
\[
y=Xw+e,
\]
where $y$ is a vector of observed values (labels), $X$ is the design
matrix with $i$-th row representing observation $x_{i}$ and defined
by a vector-valued feature map $\phi\left(x_{i}\right)$ (e.g. in
polynomial regression $\phi_{k}\left(x_{i}\right)=x_{i}^{k}$ with
$k$ ranging from $0$ to the order of a polynomial), $w$ is a vector
of model parameters and $e$ is a disturbance term (assume symmetric
uni-modal distribution of $e$ and $\mathrm{cov}\left(e\right)=\sigma^{2}I$).
Obviously, we want to find the model parameters with a minimum number
of training examples and a minimum uncertainty.

\begin{figure}[H]
\begin{centering}
\includegraphics[width=1\textwidth]{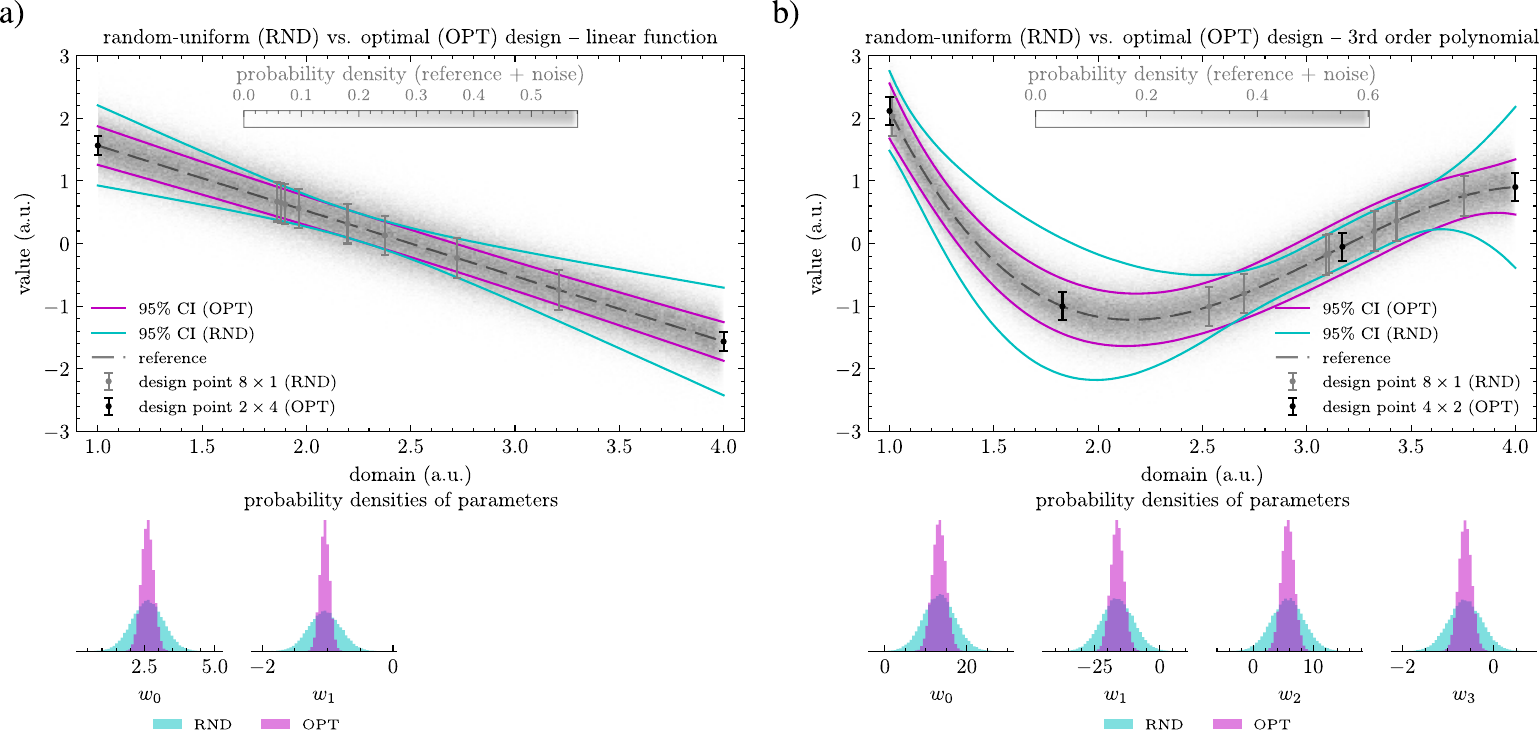}
\par\end{centering}
\caption{Comparison of optimal design (\emph{D}-optimality) for polynomial
regression with randomly selected point from a domain. Note that all
points were selected before making estimates. Confidence interval
(CI) and density of\.{ }parameters are estimated using Monte Carlo
(MC) method with $5\times10^{4}$ samples. Error bars represent design
points, i.e. points where `measurements' were made in each MC iteration.
Functions and their parameters are defined as follows: $f(x)=w_{0}x^{0}+w_{1}x^{1}+\ldots w_{N}x^{N}.$
Space-filling designs will yield similar results to the optimal design.
Note that significant difference in the performance is emphasised
by the unfavourable sample.}

\label{fig:design_example}
\end{figure}

Consider the example of \emph{D}-optimality which aims to maximise
the determinant of the Fisher information matrix (FIM). For least-square
estimators FIM is simply given as $\mathscr{I}=\sigma^{-2}X^{\top}X$.
It is related to the covariance of $w$ which is given by $\Sigma=\sigma^{2}\left(X^{\top}X\right)^{-1}$.
Hence, by maximising the information $\mathscr{I}$ we minimise the
uncertainty $\Sigma$. We see that the objective can be achieved without
referring to $y$'s at all and that we can quantify the effect of
selecting different training examples without $\sigma$, i.e. the
noise/uncertainty associated with the data. Hence, the optimal design
will be a function solely of $X$ \footnote{Additionally, if the problem is ill-conditioned, the matrix $X^{\top}X$
will have zero determinant and will not be invertible.}. An example of applying an optimal design to a simple polynomial
model can be found in figure \ref{fig:design_example}. 

In this example, we intentionally used random uniform sampling rather
than the equal division of the domain. While the latter seems like
a natural choice, when the feature space is high-dimensional, some
form of random sampling is considered to be an adequate solution.
Therefore, the results can be considered as a representative illustration
of the benefits that optimal design can provide. Furthermore, by selecting
low-order polynomials, we can show that optimal designs can differ
from the usual intuition. While after some careful consideration,
we could conclude that the best design for linear models is to concentrate
our whole budget on the edges of the domain, the $4\times2$ (four
points sampled twice) can be considered as more difficult to foresee
without formal calculations. Particularly, it can be difficult in
high-dimensional and non-Euclidiean spaces.

Note that there are other popular criteria such as the \emph{A}-optimality
-- minimise $\mathrm{tr}\Sigma$, \emph{E}-optimality -- maximise
minimal eigenvalue of $\mathscr{I}$ or \emph{G}-optimality which
seeks to minimise the maximum prediction variance \cite{Rasch_PWN_1991}.
These are just few examples and an appropriate selection will depend
on the family of models and expectations with respect to the performance
of estimators.%

\section{Example of the implementation in Python\label{sec:maxmin_python}}

In this section we present a simple implementation of the conditional
max-min design (figure \ref{fig:the_alg}) using the example of two-dimensional
Euclidean space. This will be a case of space filing design which
is well performing design overall and it is close to optimal for polynomials
of high degree or kernel based methods with Gaussian kernel. 

We start by generating the pool of candidates. Here it will be a dense
uniform grid of points covering the entire domain $\left[-1,1\right]\times\left[-1,1\right]$.
The Python code using the NumPy library, imported under the 'np' label,
can be found in the listing \ref{alg:gen_grid}.
\begin{algorithm}[H]
\begin{lstlisting}[language=Python]
x1, x2 = np.mgrid[-1:1.05:.05, -1:1.05:0.05]
x1, x2 = x1.flatten(), x2.flatten()
X = np.vstack((x1, x2)).T
rng = np.random.default_rng()
X = rng.permutation(X, axis=0)
\end{lstlisting}

\caption{Generation of uniform grid of points.\label{alg:gen_grid}}
\end{algorithm}
In the context of ML potentials, in realistic scenarios we do not
have well-defined metric, and even random uniform sampling is impossible
or impractical, let alone the importance sampling. However, such an
idealised test case allows for an easier assessment of the algorithm
and its implementation.

\begin{figure}[H]
\begin{centering}
\includegraphics[width=1\textwidth]{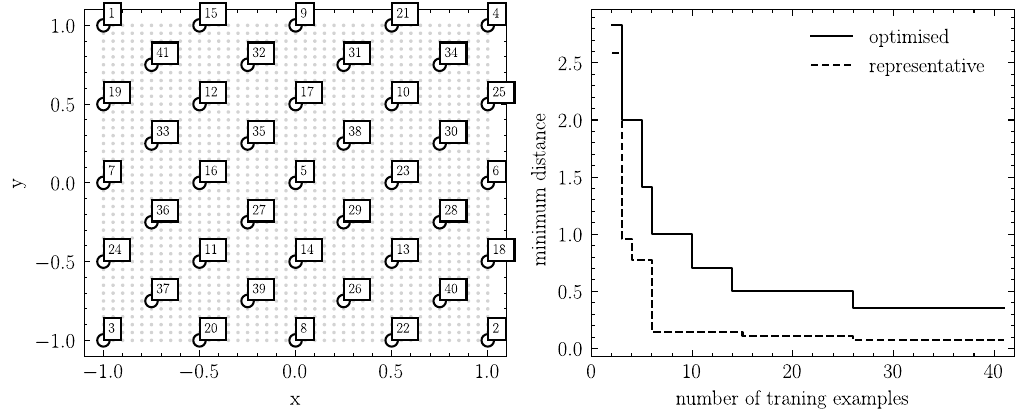}
\par\end{centering}
\caption{Illustration of the max-min design. On the left, optimal points selected
form a candidates distributed on a regular grid. Ordering is `random'.
Labels indicate in which order points were added to the design. On
the right, minimum distance of the max-min design compared to random,
indiscriminate selection of candidates.}

\label{fig:grid_desing}
\end{figure}

The main optimisation loop, presented in listing \ref{alg:opt_grid},
has a very simple implementation that takes advantage of vectorised
element-wise operations.
\begin{algorithm}[H]
\begin{lstlisting}[language=Python]
def delta(x, y, p=2):    
    return np.linalg.norm(x - y, axis=1, ord=p)


i = np.argmax(np.linalg.norm(X, axis=1))
X[[0, i]] = X[[i, 0]]
S = [0] Psi = delta(X, X[S[0]])
for k in range(40):
    Psi = np.minimum(Psi, delta(X, X[S[-1]]))
    S.append(np.argmax(Psi))
\end{lstlisting}

\caption{Selection of optimal training points. \label{alg:opt_grid}}
\end{algorithm}
The distance between elements is defined by the Euclidean distance
($L_{2}$ norm). The results of the optimisation are presented in
figure \ref{fig:grid_desing}.
Here, we will omit implementation of the code used to generate figures
and calculate the minimum distance. 

As in the example from figure \ref{fig:alg_test}, given the budget,
the algorithm tries to cover the whole space evenly and embrace the
whole domain. This is an important feature, as it allow us to adjust
the budget simply by selecting first $n$ candidates.
\end{document}